\documentclass{revtex4}

\usepackage{amsmath}
\usepackage{amssymb}
\usepackage{graphicx}

\begin{document}

\newcommand{\be}{\begin{eqnarray}}
\newcommand{\ee}{\end{eqnarray}}
\newcommand{\bes}{\begin{eqnarray*}}
\newcommand{\ees}{\end{eqnarray*}}

\newcommand{\bals}{\begin{align*}}
\newcommand{\eals}{\end{align*}}
\newcommand{\bmt}[1]{\mbox{\boldmath $#1$}}
\newcommand{\f}{\frac}
\newcommand{\edo}{\eta \downarrow 0}
\newcommand{\p}{\partial}
\newcommand{\w}{\tilde}
\newcommand{\wt}{\widetilde}
\newcommand{\ov}{\overline}
\newcommand{\un}{\underline}
\newcommand{\wh}{\widehat}
\newcommand{\ph}{\phantom}
\newcommand{\cp}{{\cal{P}}}
\newcommand{\ete}{{\texttt{e}}}
\newcommand{\eto}{{\texttt{o}}}
\newcommand{\bs}{\bar{\sigma}}
\newcommand{\mc}{\mathcal}
\newcommand{\mb}{\mathbb}
\newcommand{\vep}{\varepsilon}
\newcommand{\ep}{\epsilon}

\newtheorem{theorem}{Theorem}
\newtheorem{lemma}{Lemma}[section]
\newtheorem{corollary}{Corollary}[section]
\newtheorem{prop}{Proposition}

\title{Gauge invariant perturbations of self-similar
Lema\^{i}tre-Tolman-Bondi\\
spacetime: even parity modes with $l\geq 2$}

\author{Thomas J Waters}

\address{Department of Applied Mathematics, National University of Ireland, Galway}

\author{Brien C Nolan}

\address{School of Mathematical Sciences, Dublin City University, Dublin}

\begin{abstract}
  In this paper we consider gauge invariant linear perturbations of the
  metric and matter tensors describing the self-similar Lema\^{i}tre-Tolman-Bondi
  (timelike dust) spacetime containing a naked singularity. We decompose the angular part of the
  perturbation in terms of spherical harmonics and perform a
  Mellin transform to reduce the perturbation equations to a set of
  ordinary differential equations with singular points. We fix
  initial data so the perturbation is finite on the axis and the
  past null cone of the singularity, and follow the perturbation
  modes up to the Cauchy horizon. There we argue that certain scalars formed from the modes of the
  perturbation remain finite, indicating linear stability of the
  Cauchy horizon.
\end{abstract}
\pacs{04.20.Dw \and 04.30.-w}

\maketitle

\section{Introduction}

The naked singularities predicted in certain solutions to
Einstein's field equations pose a threat to the validity of
Penrose's Cosmic Censorship Hypothesis (CCH); indeed, the CCH
forbids the existence of naked singularities in generic
gravitational collapse. Nonetheless, certain counterexamples do
exist in which collapse results in a naked singularity. The best
known example is perhaps the Reissner-Nordstr\"{o}m solution (in the case $|charge|<|mass|$),
however other instances would include the Kerr spacetime
\cite{Wald}, spacetimes containing colliding plane waves \cite{jg}
and spacetimes featuring critical collapse \cite{livrev}. These
naked singularities suggest the possibility of information from
the singularity escaping to the external universe, resulting in a
loss of well-posedness of the field equations. Fortunately, the
Reissner-Nordstr\"{o}m solution also provides a paradigm for the
possible saviour of the CCH: perturbations in the metric and
matter tensor grow without bound when the Cauchy horizon is
approached, with the Cauchy horizon undergoing a `blue-sheet'
instability and becoming singular itself (see Chandrasekhar and
Hartle \cite{chandhartle}; see also Dafermos \cite{dafermos1,dafermos2}
for the Einstein-Maxwell-scalar field case and Poisson and Israel \cite{poisson}).
Thus the naked singularity is unstable
under linear perturbations, and these perturbations are essential
to give our spacetime the genericity on which the CCH depends; so
much so that it could well be that naked singularities are an
artifact of the high degree of symmetry of the spacetimes in which
they are typically observed.

A class of spacetimes with an additional degree of symmetry in
which naked singularities are commonly seen to occur is the class
of self-similar spherically symmetric (SSSS) spacetimes, for
example, certain classes of self-similar perfect fluid
\cite{oripiran} and dust \cite{joshi} solutions, and the
self-similar scalar field \cite{christi}. In previous work, the
authors have tested the stability of certain members of this
class. In \cite{BandT1} a scalar field was used to model a
perturbation and was allowed to impinge on the Cauchy horizon of a
SSSS spacetime whose matter tensor was unspecified save for
satisfying certain energy conditions, and in \cite{brienscalar} pointwise
bounds are found for a scalar wave impinging on the Cauchy horizon of a SSSS spacetime. In
\cite{BandT2} we considered gauge invariant metric and
matter perturbations of the self-similar null dust or Vaidya
solution. The present paper represents a continuation of this process, in
which we consider the perturbations of a more realistic and
relevant spacetime, the self-similar timelike dust solution of
Lema\^{i}tre-Tolman-Bondi (LTB). We model metric and matter
perturbations to take us away from the high degree of symmetry in
the background, and, after deriving initial conditions on the axis
and past null cone of the origin, allow the perturbations to
evolve up to the Cauchy horizon. There we see that certain scalars
built from the modes of the perturbations remain finite,
indicating that the Cauchy horizon associated with the
self-similar LBT spacetime is linearly stable and does not display
the `blue-sheet' instability seen in the Reissner-Nordstr\"{o}m
solution. The `modes' here refer to the coefficients of the state variables following a Mellin transform. We will use a commoving spatial coordinate $r$ and a similarity variable $y$ that is a time coordinate in the region between the past null cone of the naked singularity and the Cauchy horizon. Then the Mellin transform effects $G(y,r) \to g(y;s)r^{s-1}$. Then the statement above refers to the behaviour of the functions $g(y;s)$: when these satisfy conditions on the axis and along the past null cone that correspond to the presence of an initially finite perturbation, they remain finite at the Cauchy horizon. Thus we demonstrate that  a {\em necessary} condition for the linear stability of the Cauchy horizon is satisfied. The corresponding {\em sufficient} condition would entail demonstrating that the inverse Mellin transform of an initially finite perturbation remains finite. We discuss this resummation problem below.


%
%

The perturbation formalism of Gerlach and Sengupta \cite{GS1,GS2}
which we use in this work is very robust in that it can be applied
to any spherically symmetric background. Moreover, the formalism has
been tailored for the longitudinal or Regge-Wheeler gauge which
simplifies the matter perturbation terms. Thus this formalism has
been used by a number of authors in order to describe perturbations
of spherically symmetric spacetimes, among them perturbations of
critical behaviour in the massless scalar field by Frolov
\cite{Frolov2,Frolov1} and Gundlach and Mart\'{\i}n-Garc\'{\i}a
\cite{gundmartin}; perturbations of timelike dust solutions by
Harada et al \cite{haradaodddust,haradaevendust}; and perturbations
of perfect fluids by Gundlach and Mart\'{\i}n-Garc\'{\i}a
\cite{Gund,gundfluid}. These analyses (with the exception of
Frolov's) primarily rely on numerical evolution of the perturbation
equations; there is a gap in the literature with regards to analytic
or asymptotic solutions to perturbations of these spacetimes.

In broader terms, perturbations of the metric tensor can be
thought of as modelling gravitational waves, an important topic in
the current scientific community. This formalism has been used for
exactly that purpose by numerous authors such as Harada et al
\cite{haradaodddust,haradaevendust,harwave}, Sarbach and Tiglio
\cite{Sarbach}, and similar analyses by Nagar and Rezzolla
\cite{Rezzolla}. Gravitational waves manifest themselves at the
quadrupole and above, that is multipole mode number $l\geq 2$.
Therefore in this work we will consider only those modes $l\geq
2$. In addition, we restrict our analysis to the even parity
perturbations as it is in the even sector where the metric and
matter perturbations are fully coupled, thus presenting a more
substantial and interesting model. In the odd sector, the metric
and matter perturbations are coupled but only insofar as the
matter perturbation acts as a source term, and obeys a decoupled
equation that fully determines the evolution of the matter
perturbation. Furthermore, the master equation governing odd parity perturbations takes the form of a wave equation with a source term. This source term is completely and explicitly determined in terms of initial data, and does not give rise to any divergence. Then the perturbation may be dealt with {\em without} recourse to a Mellin decomposition using the methods of \cite{brienscalar} and it seems clear that no instability arises. In order to restrict the length of the present paper, we defer a complete discussion of the odd parity case to a future publication

The principal finding of this paper is that the Cauchy horizon
formed in the collapse of the self-similar Lema\^{i}tre-Tolman-Bondi
spacetime is stable under linear gauge invariant perturbations in
the metric and matter tensors, at the level of the Mellin modes as outlined above. In the next section we describe the
mathematical background to the stability analysis, namely we derive
the metric and matter tensor for the self-similar timelike dust
solution, we outline the perturbation formalism of Gerlach and
Sengupta, and we describe two important mathematical tools: the
Mellin integral transform and the generalized Frobenius theorem. In
Section III we test the mode stability of the LTB spacetime by finding
asymptotic limits for the perturbation modes on the axis and past
null cone of the origin, and under suitable initial conditions allow
the perturbation to evolve to the Cauchy horizon and beyond. We use
the conventions of Wald \cite{Wald} and set $G=c=1$.


\section{Preliminaries}

\subsection{The self-similar LTB spacetime}

The Lema\^{i}tre-Tolman-Bondi spacetime has been well studied in the
literature and we will not derive this solution here (but see for
example Harada et al. \cite{haradaevendust}), we merely give a
summary of the main points:

The LTB solution describes dust particles which move along
timelike geodesics in a spherically symmetric spacetime, and thus
has a matter tensor of the form \bes t_{\mu\nu}=\rho\, u_\mu
u_\nu, \ees where $u_\mu u^\mu=-1$. We use comoving coordinates
$t,r$ with $u^\mu=\delta^\mu_t$ and $u^\mu\nabla_\mu r=0$, and let
$R=R(t,r)$ denote the areal radius. Solving the field equations
gives (letting dot and prime denote differentiation w.r.t.\ $t$
and $r$ resp.) \be \rho=\f{1}{8\pi} \f{2m'}{R^2 R'},
\label{densitydust} \ee where $m$ denotes the Misner-Sharpe mass,
and in the marginally bound case $ R^3=\frac{9}{2}m(r)
\big[t_c(r)-t\big]^2, $ with $t_c(r)=\tfrac{2}{3}\sqrt{r^3/2m}.$
Thus once we have specified $m(r)$ (or alternatively $\rho(0,r)$)
we have completely determined all the unknowns.

From \eqref{densitydust} we see the density diverges when $R=0$,
that is when $t=t_c(r)$. This is the curvature singularity known as
the shell-focusing singularity, and we can interpret the function
$t_c(r)$ then as the time of arrival of each shell of fluid to the
singularity. Note there is an additional singularity known as the
shell-crossing singularity when $R'=0$. We will not consider this
singularity as one may extend spacetime non-uniquely through the
shell crossing singularity, see Nolan \cite{brienscs}. To rule out
the occurrence of the shell-crossing singularity we take $R'>0$ for
all $r>0$, see Nolan and Mena \cite{brienphillipe}.

Thus the line element for marginally bound timelike dust collapse is
\bes ds^2=-dt^2+R'^2 dr^2+R^2 d\Omega^2. \ees We, however, are
interested in self-similar collapse, and thus we look for a
homothetic Killing vector field $\xi^a$ which solves the equation
$\nabla_a\xi_b+\nabla_b\xi_a-2g_{ab}=0$. If
$\xi^a=\big(\alpha(t,r),\beta(t,r)\big)$, this returns four
equations, \bes \dot{\alpha}=1, \qquad
R'^2\dot{\beta}=\alpha',\qquad \beta R''+\beta'R'-R'+\alpha
\dot{R}'=0, \qquad \beta R'+\alpha\dot{R}-R=0.\ees From the first
equation we can write $\alpha=t+F_1(r)$, for arbitrary $F_1$. Since
$\beta R''+\beta'R'=(\beta R')'$, we may combine the third and
fourth equations to give $\alpha'=0$, and thus we may change the
origin of $t$ to set $\alpha=t$. The second equation therefore gives
$\beta=F_2(r)$, and we can make a coordinate transformation to set
$\beta=r$. The remaining equations are \bes t(R')\dot{}+r(R')'=0,
\qquad t\dot{R}+r R'-R=0.\ees The first of these equations is $\xi^a
\p_a R'=0$ which is solved if and only if $R'$ is a function of a
similarity variable, in this case $y=t/r$. Thus if we set $R=r
G(y)$, where $G$ is a function of the similarity variable, we have
$\p R/\p r=G-y (dG/dy)$, which is solely a function of $y$.

Thus the line element for a self-similar spherically symmetric
timelike dust will be \be ds^2=-dt^2+(G-y G\,')^2dr^2+r^2 G^2
d\Omega^2, \ee where from now on a prime denotes differentiation
w.r.t.\ $y$. We may use this metric now to generate the Einstein
tensor and examine the field equations, still using the co-moving
coordinates. The $rr$ component of the field equations is $G'^2+2 G
G''=0$. Integrating yields $G G'^2=p^2$, where $p$ is some constant.
The $tt$ component then gives \bes \rho=\f{1}{8\pi}\f{G G'^{\,
2}}{r^2 G^2(G-y G')},  \ees which is why we chose $G G'^{\,
2}=p^2\geq 0$. Finally integrating this equation and using
$R\vert_{t=0}=r$ we can solve for $G$ as \be G(y)=(1-\mu y)^{2/3},
\ee where $\mu=-\tfrac{3}{2}p$. We note that flat spacetime is
recovered by setting $\mu=0$.

There is a shell-focussing singularity therefore at $y=\mu^{-1}$.
Since $$\p R/\p r=(1-\mu y)^{2/3}\left(1+\tfrac{2}{3} \mu y(1-\mu
y)^{-1}\right),$$ we see that prior to the formation of the shell
focussing-singularity, $y<\mu^{-1} \Rightarrow 1-\mu y>0$, thus
$\partial R/\partial r>0$. This rules out the formation of
shell-crossing singularities prior to the formation of
shell-focussing singularities.

\bigskip

\begin{figure}[t!]
\includegraphics{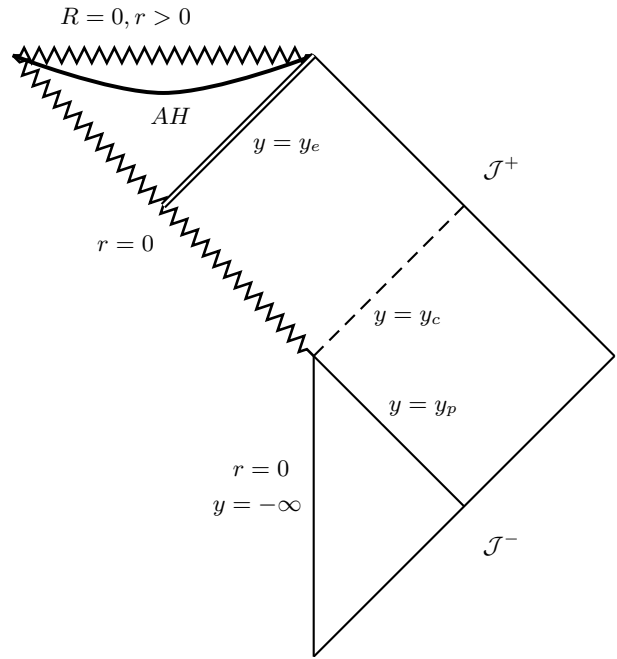} \caption{Conformal
diagram for the self-similar LTB admitting a globally naked
singularity. There are three similarity horizons at which the
similarity coordinate $y$ is null: $y=y_p$ the past null cone, $y=y_c$
shown dashed, and $y=y_e$ shown as a double line. We identify
$y=y_c$ as the Cauchy horizon, and will call $y=y_e$ the second
future similarity horizon (SFSH). The apparent horizon is shown as a
bold curve.}\label{confns}
\end{figure}

The last issue is to examine the causal structure of the
spacetime. Radial null geodesics satisfy \bes \f{dt}{dr}=\pm
\f{\partial R}{\partial r}, \ees with the plus and the minus
describing ingoing and outgoing null geodesics respectively. Since
$t=y r$ this equation may be rewritten as \bes
\f{dy}{dr}=\f{1}{r}\left(\pm \f{\partial R}{\partial r}-y\right).
\ees If there is some $y=$ constant which is a root of the right
hand side of this equation, it represents a null geodesic which
reaches the singularity in the future/past. Thus the Cauchy
horizon, $y=y_c$, is given by the first real positive zero of \be
G-y G'-y=0, \ee if one exists, and the past null cone of the
origin, $y=y_p$, is given by the root of \be G-y G'+y=0. \ee Since
$G=(1-\mu y)^{2/3}$, we find there is a Cauchy horizon, and
therefore a naked singularity, if $\mu$ is in the range \bes 0 <
\mu \leq \mu_*, \qquad \mu_*=\frac{3}{2}\left(104-60\sqrt{3}\right)^{1/3}\approx 0.638014. \ees  Moreover, when
$\mu$ is in this range, we have the following: there is one past
null cone of the origin $y_p$; there is an additional future
similarity horizon at $y=y_e>y_c$; as $\mu\to\mu_*$, $y_e\to y_c$;
and as $\mu\to 0$, $y_p\to -1, y_c \to 1$ and $y_e \to \infty$.

Thus when $0<\mu<\mu_*$, we have a spacetime with the structure
given in Figure \ref{confns}. The scaling origin at which the
singularity initially forms is the point $(t,r)=(0,0)$. The apparent
horizon forms when $g^{ab}\nabla_a R\nabla_b R=0$ which is
equivalent to $dG/dy=1$. This occurs at
$y=\tfrac{1}{\mu}\left(1-\big(\tfrac{2 \mu}{3}\big)^3\right)$, that
is, \emph{before} the formation of the shell-focusing singularity at
$y=\tfrac{1}{\mu}$.


\subsection{Gauge invariant perturbations}

We will use the formalism of Gerlach and Sengupta \cite{GS1,GS2}.
This formalism has been well used in the literature and so we will
only give an outline here for completeness (but see Gundlach and
Mart\'{i}n-Garc\'{i}a \cite{Gund} or the authors \cite{BandT2} for
a more detailed description).

We perform a 2+2 split of spacetime into a manifold spanned by
coordinates $x^A=(t,r)$ denoted $(\mc{M}^2,g_{AB})$, crossed with
unit two spheres spanned by $x^a=(\theta,\phi)$ coordinates and
denoted $(\mc{S}^2,\gamma_{ab})$. A spherically symmetric
spacetime will therefore have a metric and matter tensor given by
\bes g_{\mu\nu}dx^\mu dx^\nu = g_{AB}(x^C)
dx^A dx^B+R^2(x^C) \gamma_{ab} dx^a dx^b,\quad \ \\
t_{\mu\nu}dx^\mu dx^\nu = t_{AB}(x^C) dx^A dx^B+\tfrac{1}{2}
t^c_{\ph{a}c}\, R^2(x^C) \gamma_{ab} dx^a dx^b. \ees

Capital Latin indices will denote coordinates on $\mc{M}^2$,
lowercase Latin indices will denote coordinates on $\mc{S}^2$, and
Greek indices the 4-dimensional spacetime (i.e.\ $x^\mu=(x^A,x^a)$).
$R$ is a function on $\mc{M}^2$ and gives the areal radius.
Covariant derivatives on $\mc{M},\mc{M}^2$ and $\mc{S}^2$ are
respectively denoted \bes g_{\mu\nu;\lambda}=0, \quad g_{AB|C}=0,
\quad g_{ab:c}=0, \ees and a comma defines a partial derivative.

We write a non-spherical metric and matter perturbation as \bes
g_{\mu \nu}=\w{g}_{\mu \nu}+h_{\mu \nu}(t,r,\theta,\phi),\qquad
t_{\mu\nu}=\w{t}_{\mu\nu}+\Delta t_{\mu\nu}(t,r,\theta,\phi), \ees
where from now on an over-tilde denotes background quantities. The
spherical harmonics form a basis for functions, and from the
spherical harmonics we can construct bases for vectors, \be
\Big\{\,\, Y_{,a} \quad ; \quad S_a\equiv \epsilon_a^{\phantom{a} b}
Y_{,b} \Big\} \label{vecbases} \ee and tensors, \be \Big\{\,\,
Y\gamma_{ab} \quad ; \quad Z_{ab}\equiv
Y_{,a:b}+\f{1}{2}\,l(l+1)Y\gamma_{ab} \quad ; \quad S_{(a:b)} \Big\}
\label{tenbases} \ee where we have suppressed the mode numbers
$l,m$, $X_{(ab)}=\tfrac{1}{2}(X_{ab}+X_{ba})$ is the symmetric part
of a tensor, and $\epsilon_{ab}$ is the anti-symmetric pseudo-tensor
with respect to $\mc{S}^2$ such that $\epsilon_{ab:c}=0$. Using
these, we decompose the perturbation in terms of scalar, vector and
tensor objects defined on $\mc{M}^2$, times scalar, vector and
tensor bases defined on $\mc{S}^2$.

We write the metric and matter perturbation as \bes h_{\mu
\nu}=\left(
\begin{tabular}{cc} $h_{AB} Y$ & $h_A Y_{,a}$ \\ \emph{Symm}
& $R^2(K Y \gamma_{ab}+G Z_{ab})$ \end{tabular} \right), \quad
\Delta t_{\mu \nu}=\left(
\begin{tabular}{cc} $\Delta t_{AB} Y$ & $\Delta t_A Y_{,a}$ \\
\emph{Symm} & $R^2 \Delta t^1 Y \gamma_{ab}+\Delta t^2 Z_{ab}$
\end{tabular} \right), \label{pertev}
\ees where, as mentioned previously, we confine our interest to
even parity perturbations; that is, those defined using bases $Y$,
$Y_{,a}$, $Y\gamma_{ab}$ and $Z_{ab}$. From these, we construct a
set of gauge invariant scalars, vectors and tensors given by
\begin{subequations}\label{giseven}\be\left.\begin{array}{rcl}
k_{AB}&=&h_{AB}-(p_{A|B}+p_{B|A})
\\ k&=&K-2 v^A p_A \end{array}\right\}(metric) \\
\left.\begin{array}{rcl} T_{AB}&=&\Delta
t_{AB}-\w{t}_{AB|C}p^C-\w{t}_A^{\phantom{A}C}p_{C|B}-\w{t}_B^{\phantom{B}C}p_{C|A}
\\ T_A&=&\Delta t_A-\w{t}_A^{\phantom{A}C}p_{C}-R^2(\w{t}^a_{\phantom{a}a}/4)G,_A \\ T^1&=& \Delta t^1
-(p^C/r^2)(R^2
\w{t}^a_{\phantom{a}a}/2),_C+l(l+1)(\w{t}^a_{\phantom{a}a}/4)G \\
T^2&=& \Delta t^2-(R^2 \w{t}^a_{\phantom{a}a}/2)G
\end{array}\right\}(matter) \ee\end{subequations}
where $p_A=h_A-\f{1}{2}R^2 G_{,A}$, and $v_A=R,_A/R$.

We may then recast the perturbation equations entirely in terms of
these gauge invariant quantities (see Appendix A). Finally we must
consider what to measure on the relevant surfaces to test for
stability. As explained in \cite{BandT2}, our `master' function
will be \be \delta P_{-1}=|\delta \Psi_0 \delta \Psi_4|^{1/2}, \ee
where \be \delta \Psi_0=\f{Q_0}{2r^2}\w{\ell}^A \w{\ell}^B k_{AB}
, \quad \delta \Psi_4=\f{Q_0^*}{2r^2} \w{n}^A \w{n}^B k_{AB} , \ee
with \bes Q_0 = \w{w}^a \w{w}^b Y_{:ab}, \ees where
$\w{\ell}^\mu,\w{n}^\mu,\w{m}^\mu=r^{-1} \,
\w{w}^\mu(\theta,\phi)$ and $\w{m}^{*\mu}$ are a null tetrad of
the background and the $*$ represents complex conjugation. We note
that $\delta P_{-1}$ is a fully gauge invariant scalar, being both identification and tetrad gauge invariant (see \cite{BandT2}).


\subsection{The Mellin transform}

The Mellin transform is an integral transform related to the Laplace
transform and is particularly useful for equations deriving from
self-similar spacetimes. It is defined by \be
G(y;s)=\mathbb{M}[g(x,r)](r\to s):=\int_0^\infty g(y,r) r^{s-1}dr
\ee with $s\in\mathbb{C}$. For this transform to exist, there will
be a restriction on the allowed values of $s$, typically to lie in a
strip in the complex plane with $\sigma_1<Re(s)<\sigma_2$. The
inverse Mellin transform is given by \be
g(y,r)=\mathbb{M}^{-1}[G(y;s)]=\f{1}{2 \pi i} \int_{c-i \infty}^{c+i
\infty} r^{s} G(y;s) ds, \ee where $c \in \mathbb{R}$ is such that
$\sigma_1<c<\sigma_2$. To recover the original function from the
Mellin transform, we integrate over the vertical contour in the
complex plane of $s$ given by $Re(s)=c$. We emphasize, as this will
be crucial later, that we do not integrate over all values of $s$ in
the interval $\sigma_1<Re(s)<\sigma_2$, only over the vertical
contour defined by a specific value of $Re(s)$ in this interval,
which we are free to choose.

The perturbation equations will reduce to systems of ode's in the
individual modes of the Mellin transformed variables. Resumming the
modes to recover the original function is an extremely complicated
task and is beyond the scope of this paper (although see
\cite{BandT2} for a discussion), however we will point out that
although the finiteness of each mode is not a sufficient condition
for the finiteness of the resummed original function, it is a
necessary condition. Thus we will adopt the following minimum
stability requirement: for the inverse Mellin transform to exist on
a surface we must have each mode of the Mellin transformed quantity
finite on that surface.


\subsection{Extension to the Frobenius theorem}

The theorem of Frobenius is particularly useful in finding power
series solutions to ordinary differential equations near regular
singular points. Consider the following $n$th order ODE in $f(x)$;
there is a regular singular point at $x=0$ if the ODE is of the
form \be x^n f^{(n)}(x)+x^{n-1} b_1(x) f^{(n-1)}(x)+ \ldots +
b_{n}(x) f(x)=0, \label{ndrsp}\ee with each $b_i$ analytic at
$x=0$. We can Taylor expand each $b_i$ about $x=0$, and we denote
such as expansion as \bes b_i(x) = \sum_{m=0}^\infty b_{i,m}\,
x^m. \ees The so-called indicial exponents (see below) determine
the leading behaviour of the series solutions. It is well known
that when the indicial exponents repeat the solution must contain
a logarithmic term, and when they differ by integers the solution
may or may not contain a logarithmic term (see, for example,
\cite{Wang}). To clarify the structure of the solution
when the roots differ by integers we give the following theorem
due to Littlefield and Desai \cite{LittlefieldDesai}.

\begin{theorem} \label{LittleDes} \emph{($n$th order Frobenius
theorem)}\\ Let f(x) solve an ODE of the form \eqref{ndrsp}. Then
the indicial equation is \bes
I_n(\lambda)\equiv\lambda(\lambda-1)\ldots(\lambda-n+1)+b_{1,0}
\lambda(\lambda-1)\ldots(\lambda-n+2)+\cdots+b_{n-1,0}
\lambda+b_{n,0}, \ees whose roots are the indicial exponents. Collect together the indicial exponents which differ by integers into groups, and
order the elements of each group as \bes \{\lambda_1, \lambda_2,
\ldots, \lambda_j, \ldots \} \ees such that
$\lambda_i>\lambda_{i+1}$. Then the solution corresponding to $\lambda_1$ is $f_1(x)=\sum_{m=0}^\infty A_m x^{m+\lambda_1}$, and a linearly independent solution
corresponding to $\lambda_j$ is \bes f_j(x) = K_1 \log^{j-1}x\sum_{m=0}^\infty A_m x^{m+\lambda_1}+ \sum_{i=2}^j\left(
\beta_i \, K_i \, \log^{(j-i)}x \, \sum_{m=0}^\infty
\f{\p^{(i-1)}}{\p \lambda^{(i-1)}}
\Big[(\lambda-\lambda_i)A_m\Big]_{\lambda_i} \, x^{m+\lambda_i}
\right) \ees where \bes K_i = \lim_{\lambda \rightarrow \lambda_j}
\left( \f{\p^{(i-1)}}{\p \lambda^{(i-1)}} \left[
(\lambda-\lambda_j)^{(j-1)} \f{A_{\delta_i}}{A_0} \right] \right),
\quad K_j=1 \ees and \bes \delta_i = \lambda_i-\lambda_j \in
\mathbb{N}, \qquad \beta_i = \f{(j-1)(j-2)\ldots(j-i+1)}{i-1}, \quad
\beta_1=\beta_j=1. \ees
\end{theorem}


\section{Perturbations of self-similar LTB spacetime}

We will consider only modes $l\geq 2$. We assume the perturbed
matter tensor remains that of dust, \bes
\w{t}_{\mu\nu}+\Delta t_{\mu\nu}=(\w{\rho}+\delta
\rho)(\w{u}_\mu+\delta u_\mu)(\w{u}_\nu+\delta u_\nu). \ees If we
write the angular part in terms of the spherical harmonics, $\delta
\rho=\varrho Y$ and $\delta u_\mu=\zeta_A Y+\zeta Y_{,a}$, then for
$\w{u}_\mu+\delta u_\mu$ to be a unit, future pointing, timelike
geodesic of the perturbed spacetime, as the conservation equation
$\nabla^\mu \,t_{\mu\nu}=0$ implies must be the case, we must have
$\delta u_\mu=\gamma_{,A}Y+\gamma Y_{,a}$ for some scalar $\gamma$.
Additional, we must have \be h_{tt}=-2\gamma_{,t}. \label{addit} \ee


Using the Regge-Wheeler gauge (in which $h_A=G=0=p_A$), we may write
the gauge invariant
matter objects as \bes T_{AB}=\left(%
\begin{array}{cc}
  \varrho+2\w{\rho}\gamma_{,t} & \w{\rho}\gamma_{,r} \\
  \w{\rho}\gamma_{,r} & 0 \\
\end{array}%
\right), \quad T_A=\left(%
\begin{array}{c}
  \w{\rho}\gamma \\
  0 \\
\end{array}%
\right), \quad T^1=T^2=0. \ees Next we calculate the full set of
perturbed field equations as given in \eqref{perteqns}. We use the
equation $k_A^{\ph{A}A}=0$ to remove $k_{rr}=R'^2 k_{tt}$, and we
use the $tt$ component of \eqref{tenspert} to define $\varrho$ in
terms of the other perturbation variables.

Thus we have a set of five second order, coupled, linear, partial
differential equations in the four unknowns $\{k_{tt}, k_{tr}, k,
\gamma\}$, and the two dependent variables $t,r$. (Had we removed
$k_{tt}$ with $k_{tt}=-2\gamma_{,t}$ \eqref{addit}, these would be
third order in $\gamma$.) In this set of
equations we make the
change of coordinates \be (t,r)\to(y=\tfrac{t}{r},r),\ee and then
perform a Mellin transform over $r$, reducing the problem to five
second order ordinary differential equations in $y$, the similarity
variable, and parameterised by $s$, the transform parameter. The
Mellin transforms of our unknowns can be written \bes k_{tt}=r^s
A(y;s), \quad k_{tr}=r^s B(y;s), \quad k=r^s K(y;s), \quad
\gamma=r^{s+1} H(y;s), \ees thus the four unknowns of our set of
ODE's are $\{A,B,K,H\}$.

The future pointing ingoing and outgoing radial null geodesic
tangents of the background spacetime in $t,r$ coordinates are \bes
\w{\ell}^A=\f{1}{\sqrt{2}R'}\left(R',1\right), \quad
\w{n}^A=\f{-1}{\sqrt{2}R'}\left(-R',1\right) \ees respectively,
since we restrict $R'>0$ to avoid the shell crossing singularity
occurring before the shell-focussing singularity. The $\mc{M}^2$
portion (i.e.\ neglecting the angular part) of the perturbed Weyl
scalars then becomes \bes
\delta\Psi_0=\f{1}{r^2}\left(k_{tt}+\f{k_{tr}}{R'}\right), \qquad
\delta\Psi_4=\f{1}{r^2}\left(k_{tt}-\f{k_{tr}}{R'}\right). \ees
After a change of coordinates, and Mellin transform, we may write
each mode of these scalars in terms of $A$ and $B$. We define a new
variable $D=A+B/(G-yG')$, and the scalars' modes simplify to \bes
\delta\Psi_0= r^{s-2}D, \qquad \delta\Psi_4= r^{s-2} (2A-D),\ees and
$\delta P_{-1}=|\delta\Psi_0 \delta\Psi_4|^{1/2}$. We must find
solutions to the set of ODE's and use them to evaluate these modes
on the relevant surfaces.

\bigskip
\vspace{-2pt} We can write this set of second order ODE's as a first
order linear system \be Y'=M(y)Y \label{linsys} \ee where a prime denotes
differentiation w.r.t.\ $y$, and $Y=(A,D,K,H)^T$. We note that one
of the equations in the system is $H'=-A/2$, and thus we have
recovered \eqref{addit}, since $\p/\p t=\tfrac{1}{r}\p/\p y$. Due to
its length, we give the components of the matrix $M$ in Appendix 2.

Examining the leading order coefficient matrix near the axis reveals
that the axis corresponds to an irregular singular point of the system \eqref{linsys}, with multiple zero
eigenvalues, and a number of off-diagonal entries in its Jordan
normal form; all of which make the system methods used in
\cite{BandT2} very unattractive. In any case, we anticipate that the
system methods would break down when eigenvalues of leading matrices
differ by integers, suggesting we would at some stage need to
decouple an equation in one variable, and use its solution as an
inhomogeneous term to integrate the other equations. We will sketch
the decoupling of an equation in $H$.

This system can be written as four first order equations, \bes
h_1(A,D,K,H,A')=0, \;\; h_2(A,D,K,H,D')=0, \;\;
h_3(A,D,K,H,K')=0,\;\; h_4(A,H')=0.\ees We solve the first equation
for $D=f_1(A,K,H,A')$ and substitute this into the other three
equations, giving \bes h_5(A,K,H,A',K',H',A'')=0, \quad
h_6(A,K,H,A',K')=0,\quad h_4(A,H')=0. \ees Combining $h_5$ and $h_6$
to remove $K'$ means we can solve for $K=f_2(A,H,A',H',A'')$, and we
are left with two equations, \bes h_7(A,H,A',H',A'',H'',A''')=0,
\quad h_4(A,H')=0.\ees Finally we remove $A$ for a fourth order ODE
in $H$, \bes h_0(H,H',H'',H''',H'''')=0. \ees

The other variables can be calculated when the solutions for $H$ are
found, as \bes A=g_1(H'), \quad D=g_2(H,H',H'',H'''), \quad
K=g_3(H,H',H'',H'''), \ees and thus we can write the scalars
$\delta\Psi_{0,4}$ in terms of $H$ and its derivatives.

Having derived the necessary equations, we pause to outline our general strategy for studying linear stability. The perturbation equations comprise a first order system of equations in certain perturbation variables $X(y,r)$ from which we can extract physically significant quantities. Of principal importance here is $\delta P_{-1}$. We seek to impose initial and boundary conditions on the perturbations that correspond to the most general, initially finite perturbation that satisfies appropriate conditions at the axis. This perturbation is allowed to evolve up to the Cauchy horizon, and we then try to determine if the perturbation has remained finite. The question of what is meant by 'finite' is important. A minimal condition is that $\delta P_{-1}$ be bounded on the past null cone. This however would allow for an infinite energy content over the past null cone - or over a space-like surface an arbitrarily short time to the future of the past null cone. This leads naturally to the consideration of the $L^2$ norm of the perturbation. This is finite if and only if (by Plancherel's theorem) the $L^2$ norm of the Fourier transform of the perturbation is finite. But the Fourier transform is related to the Mellin transform by the complex rotation $s\to iz$ and $r=\ln \rho$. So we are led naturally to consider finiteness of the Mellin transform (a weaker condition than finiteness of the $L^2$ norm) as a minimal condition for finiteness of the perturbation. This is the condition on which we will focus: the perturbation $X(y,r)$ will be referred to as finite at time $y_0$ if the Mellin transform $x(y;s)$ is finite at time $y=y_0$.

\subsection{Axis.}

We consider first the axis, $y=-\infty$. We make the transformation
$y=-1/w$ to put the axis at $w=0$, and then the transformation
$w=z^3$ to ensure integer exponents in the series expansions about
the axis of the coefficients of the differential equation. We find
$z=0$ is a regular singular point of the fourth order ODE in $H$,
which we will write as \be
\sum_{j=0}^4 z^j\big[h_j+O(z)\big]\f{d^j H}{dz^j}=0, \qquad h_j \neq
0. \ee The indicial exponents near $z=0$ are
 \be \{-3, 2, -4-l-3s, -3+l-3s\}. \label{evalsh}
\ee  The ambiguity of the value of $s$ complicates matters regarding
the position of logarithmic terms in the full solution, so we will
begin by reminding ourselves of the two conditions the solutions
must solve:
\begin{enumerate} \item The solution must exist; that is we must
be able to recover the original function from its Mellin transform.
Our minimum stability requirement for this to hold is that an
acceptable solution is one which does not diverge on the relevant
surface. \item The solution must be such that each mode of
$\delta\Psi_{0,4}$ is finite on the axis.
\end{enumerate}


Consider the indicial exponent $-3$. Regardless of the values of the
other exponents, the corresponding solution will contain at least
the series $\sum_{m=0}^{\infty} A_m z^{m-3}$. This is certainly not
convergent, it diverges at $z=0$ ($A_0\neq 0$). Thus we must not
consider this solution.


In examining requirement 2, we expand the coefficients of $H$ and
its derivatives in $\delta\Psi_{0,4}$ around $z=0$, and we find the
dominant term is \bes \delta\Psi_{0,4} \sim r^{s-2} z^4 H'. \ees
Consider the indicial exponent $l-3s-3$. The contribution to the
general solution corresponding to this eigenvalue is  \bes (\textrm{Logarithmic
terms})\times(\textrm{Series})\,+\sum_{m=0}^\infty A_m z^{m+l-3s-3},
\ees where the first portion of this solution depends on the other
eigenvalues, and may not even be present. Near $z=0$, we find
$\delta\Psi_{0,4}\sim z^{l-6}$ due to the second term. We would
certainly expect these scalars to be finite on the axis for the
quadrupole and other modes with $l<6$, thus we must rule out this
solution.

Similarly for the indicial exponent $-4-l-3s$, we find
$\delta\Psi_{0,4}\sim z^{-l-7}$ near $z=0$. Thus we must also rule
out this solution.

Finally for indicial exponent $2$, we see the solution \be H=
\sum_{m=0}^\infty A_m z^{m+2} \ee is convergent (near $z=0$)
$\forall\ s$, and thus satisfies our minimum stability requirement.
Further, the scalars $\delta\Psi_{0,4}$ will be finite on the axis
for $Re(s)\geq 1/3$. Thus we have found a one-parameter family of
solutions near the axis.

\subsection{Past null cone}

The past null cone, $y=y_p$, is the real, negative root of $G-y
G'+y=0$, where $G(y)=(1-\mu y)^{2/3}$. There is only one real root (when $0<\mu<\mu_*$), and it is
parameterized by $\mu$. Thus we may write \bes G-y G'+y=(y-y_p)
F(y), \qquad F(y_p)\neq 0. \ees

$y_p$ is a very cumbersome surd, and is quite difficult to work
with. Instead, we draw out the nature of the coefficients of the
$H$-equation by using $G'(y_p)=(G(y_p)+y_p)/y_p$. We find that
setting $G'=(G+y)/y$ makes each coefficient vanish, except for the
coefficient of the highest derivative. Thus we may write the
$H$-equation as \bes (G-y
G'+y)[m_0+O(y-y_p)]H^{(4)}+[n_0+O(y-y_p)]H^{(3)}+[p_0+O(y-y_p)]H^{(2)}&
\\+[q_0+O(y-y_p)]H'+[r_0+O(y-y_p)]H&=0,\ees where $m_0,n_0$ etc.\ are
the first nonzero terms in the series expansions about the past null
cone.

We may write this in canonical form as \bes (y-y_p)^4
H^{(4)}+(y-y_p)^3 \,b_1(y) H^{(3)}+(y-y_p)^2\,b_2(y) H''+\ldots=0.
\ees If the series expansions of the $b_i$ about the past null cone
are denoted $b_i=\sum_{j=0}^{\infty}b_{i,j}\,(y-y_p)^j$, then the
first few terms in the expansions of the $b_i$ about $y=y_p$ are \be
\begin{array}{lllll}
  b_{1,0}=n_0/(m_0 F_p) & b_{1,1}=\ldots &  &  &  \\
  b_{2,0}=0 & b_{2,1}=p_0/(m_0 F_p) & b_{2,2}=\ldots &  &  \\
  b_{3,0}=0 & b_{3,1}=0 & b_{3,2}=q_0/(m_0 F_p) & b_{3,3}=\ldots &  \\
  b_{4,0}=0 & b_{4,1}=0 & b_{4,2}=0 & b_{4,3}=r_0/(m_0 F_p) & b_{4,4}=\ldots \\
\end{array}\label{coeffexp} \ee where $F_p=F(y_p)$. Therefore $y=y_p$ is a regular singular point of this ordinary
differential equation, and the indicial equation for a fourth order
ODE is \bes
\lambda(\lambda-1)(\lambda-2)(\lambda-3)+b_{1,0}\lambda(\lambda-1)(\lambda-2)+b_{2,0}\lambda(\lambda-1)
+b_{3,0}\lambda+b_{4,0}=0. \ees Thus the indicial exponents are \bes
\big\{\ 0,1,2,3-b_{1,0}\equiv\sigma\ \big\}. \ees To determine what
exactly $\sigma$ is, we note \bes F_p=F(y_p)=\lim_{y\to y_p}
\f{G-yG'+y}{y-y_p}=1-y_p G''(y_p), \ees using l'H\^opital's rule,
and thus \be
\sigma=3-\left[\f{7-s+2\left(\f{y}{G}+\f{G}{y}\right)}{1-y
G''}\right]_{y=y_p}. \label{sigpnc}\ee We note that $\sigma=s$ in
the limit $\mu\to 0$.

We may find the solutions due to these indicial exponents from the
analysis in \S II.D. Let us consider first the case
$\sigma\notin\mathbb{Z}$. We group together the indicial exponents as \bes \big\{2,1,0\big\},\quad\big\{\sigma\big\} \ees since $\sigma\notin\mathbb{Z}$. Then, according to Theorem \ref{LittleDes}, the general solution has the form \begin{align} H|_{y=y_p}=&\ h_1 \left[\sum_{m=0}^\infty A_m (y-y_p)^{m+2}\right]+h_2\left[K_1 \log x \sum_{m=0}^\infty A_m(y-y_p)^{m+2}+\sum_{m=0}^\infty B_m (y-y_p)^{m+1}\right] \nonumber \\ &+h_3\left[\bar{K}_1 \log^2 x \sum_{m=0}^\infty A_m(y-y_p)^{m+2}+\bar{K}_2\beta_2\log x\sum_{m=0}^\infty B_m (y-y_p)^{m+1}+\sum_{m=0}^\infty C_m (y-y_p)^{m}\right] \nonumber \\ &+h_4\left[\sum_{m=0}^\infty D_m(y-y_p)^{m+\sigma}\right], \end{align} where the $h_i$ are constants of integration and we have used an overbar to distinguish the $K$ coefficients in the second and third solution.

We see the general solution contains three logarithmic
terms, each multiplied by a constant. For the fourth order ODE in
$H$ we are considering here, these constants are (again from Theorem \ref{LittleDes}) \begin{align*}
&K_1=\lim_{\lambda\to 1}
\big[(\lambda-1)A_1(\lambda)\big]=\f{(b_{3,1}+b_{4,1})}{(2b_{1,0}-2)},\\
 &\bar{K}_1=\lim_{\lambda\to
0}\big[\lambda^2
A_2\big]=\f{b_{4,1}(b_{3,1}+b_{4,1})}{(2-b_{1,0})(2b_{1,0}-2)},\qquad  \bar{K}_2=\lim_{\lambda\to 0}\left[\f{d}{d\lambda}(\lambda^2
A_1)\right]=\f{b_{4,1}}{(2-b_{1,0})},
\end{align*} where we have set $A_0=1$. Crucially, since [\eqref{coeffexp}]
$b_{3,1}=b_{4,1}=0$, each of these terms vanish, and thus when
$\sigma \notin \mathbb{Z}$, we have a general solution \be
\hspace{-12pt} H|_{y=y_p}=h_1\sum_{m=0}^\infty A_m
(y-y_p)^{m+2}+h_2\sum_{m=0}^\infty B_m
(y-y_p)^{m+1}+h_3\sum_{m=0}^\infty C_m
(y-y_p)^{m}+h_4\sum_{m=0}^\infty D_m
(y-y_p)^{m+\sigma},\label{pncltb}\ee with each series linearly
independent. Our minimum stability requirement for these solutions
will be satisfied for $Re(\sigma)>0$.

Now we examine the scalars $\delta\Psi_{0,4}$ near the past null
cone, and we find we can write \be \delta\Psi_0\sim c_1H+c_2H'+c_3
H''+c_4(y-y_p)H^{(3)}, \label{scpnc} \ee with a similar expression
for $\delta\Psi_4$. The scalars are automatically finite on $y=y_p$
for the first three series in \eqref{pncltb}. For the fourth series,
we find surprisingly that $c_3
\sigma(\sigma-1)+c_4\sigma(\sigma-1)(\sigma-2)=0$ for both
$\delta\Psi_0$ and $\delta\Psi_4$; that is the coefficient of the
leading term, which goes like $(y-y_p)^{\sigma-2}$, vanishes
exactly. Thus for finite scalars on the past null cone due to the
fourth solution, we require only $Re(\sigma)>1$.

Now let us consider $\sigma\in\mathbb{Z}$. Firstly if $\sigma<0$,
the minimum stability requirement is not met and we certainly cannot
recover $\gamma$ from $H$ via the inverse Mellin transform; thus we
consider $\sigma\geq 0$. Now we note an important point regarding
the Frobenius method: if two indicial exponents differ by an
integer, the solution corresponding to the lowest index may contain
a logarithmic term; however if two indicial exponents are equal, the
second solution \emph{must} contain a logarithmic term.

If $\sigma=0$, then there will be a solution which has leading term
$\ln(y-y_p)$, which diverges at the past null cone, and thus the
minimum stability requirement is not satisfied. If $\sigma=1$, the
corresponding leading term is $(y-y_p)\ln(y-y_p)$, which is finite
in the limit $y\to y_p$. Thus we only consider $\sigma>0$, when
$\sigma\in\mathbb{Z}$.

When calculating the scalars $\delta\Psi_{0,4}$, we see from
\eqref{scpnc} that if $\sigma=1$, then $H'\sim\ln(y-y_p)$, and thus
we must discount $\sigma=1$. Again, when $\sigma=2$, we find
$H''\sim \ln(y-y_p)$, however for $\sigma\geq 3$ we have
$\delta\Psi_{0,4}\sim O(1)$.

Thus for the scalars to be finite on the past null cone, we require
$s$ to be such that $Re(\sigma)>1$, with the exception of
$\sigma=2$.

\begin{figure}[t]
\begin{center}
\includegraphics{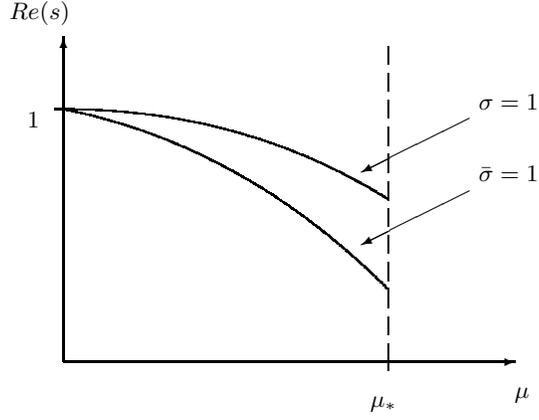}
\end{center} 
\caption{The lines $\sigma=1$ and $\bar{\sigma}=1$ plotted in the
 $Re(s),\mu$ parameter space for $0<\mu<\mu_*$.}
\label{notintltb}
\end{figure}

\subsection{Cauchy Horizon}

The Cauchy horizon, denoted by $y=y_c$, is the first real root of $G-yG'-y=0$ where $G=(1-\mu y)^{2/3}$ and, as described in \S II.A, exists and is unique for $0<\mu<\mu_*$. The situation on the Cauchy horizon is very similar to the past null
cone: we obtain a fourth order ODE in $H$ with $y=y_c$ as a regular singular
point, use series expansions about $y=y_c$ of the coefficients of the
differential equation in the form \eqref{coeffexp}, and find indicial
exponents $\{0,1,2,\bar{\sigma}\}$ where \be
\bar{\sigma}=3-\left[\f{7-s-2\left(\f{y}{G}+\f{G}{y}\right)}{1+y
G''}\right]_{y=y_c}, \qquad \lim_{\mu\to 0}\bar{\sigma}=s.
\label{sigch}\ee

When $\bar{\sigma}\notin\mathbb{Z}$, all the logarithmic terms in
the general solution vanish as at the past null cone. The scalar
$\delta\Psi_{0}$ can be written near $y=y_c$ as \bes \delta\Psi_0
\sim \bar{c}_1 H+\bar{c}_2 H'+\bar{c}_3 H''+\bar{c}_4(y-y_c)
H^{(3)}, \ees with a similar expansion for $\delta\Psi_4$. Again,
the coefficient of the leading term due to the solution due to the
indicial exponent $\bar{\sigma}$ vanishes, and we find the scalars
will be finite on the Cauchy horizon iff $Re(\bar{\sigma})>1,$ when
$\bar{\sigma}\notin\mathbb{Z}$.

When $\bar{\sigma}\in\mathbb{Z}$, we find, for the same reasons as
at the past null cone, we must rule out $\bar{\sigma}\leq 1$; when
$\bar{\sigma}=2$ the scalars diverge like $\ln(y-y_c)$; and when
$\bar{\sigma}\geq 3$ the scalars are finite on the Cauchy horizon.


Let us consider first the clearer picture, when neither $\sigma$ or
$\bar{\sigma}$ are integers. Both $\sigma$ and $\bar{\sigma}$ are
parameterized by $s$ and $\mu$, and thus we can plot the line in the
$Re(s),\mu$ parameter space where $\sigma=1$ and $\bar{\sigma}=1$.
We give this schematically in Figure \ref{notintltb} for
$0<\mu<\mu_*$.

We interpret this plot so: for every $\mu$, if $Re(s)$ is such that
the point $(\mu,Re(s))$ is above the line $\sigma=1$, the scalars
will be finite on the past null cone. Similarly, if $Re(s)$ is such
that the point $(Re(s),\mu)$ is above the line $\bar{\sigma}=1$, the
scalars will be finite on the Cauchy horizon. As the
$\bar{\sigma}=1$ line is always below the $\sigma=1$ line for
$0<\mu<\mu_*$, this means that \emph{all} perturbations which are
finite on the past null cone at the level of the modes of the Mellin transform will be finite on the Cauchy horizon at the same level,
when $\sigma,\bar{\sigma}\notin\mathbb{Z}$. It remains to consider the problem of resummation; this is discussed below.

\bigskip

\begin{figure}[t]
\begin{center}
\includegraphics[width=0.35\textwidth]{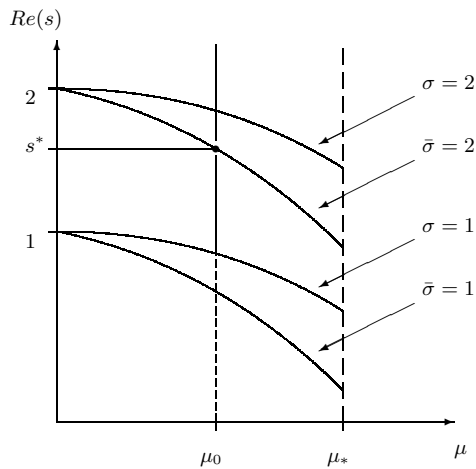}
\end{center} 
\caption{The lines $\sigma=1,2$ and
$\bar{\sigma}=1,2$ plotted in the
 $Re(s),\mu$ parameter space for $0<\mu<\mu_*$.}
\label{intltb}
\end{figure}

When $\sigma,\bar{\sigma}\in\mathbb{Z}$, the picture is a touch more
intricate, due to the fact that $\sigma=2$ or $\bar{\sigma}=2$ will
give a divergence in the scalars. Consider Figure \ref{intltb}, and
let's choose a particular value for $\mu$, $\mu_0$ where
$0<\mu_0<\mu_*$. The solid portion of the line $\mu=\mu_0$
represents all the allowable values (from the point of view of
initial data) of $Re(s)$ for this $\mu_0$, with the exception of
where the line intersects $\sigma=2$. We see that this line must
intersect $\bar{\sigma}=2$ at some point $(\mu_0,s^*)$, represented
by the black dot in Figure \ref{intltb}.

\begin{figure}[t]
\begin{center}
\includegraphics{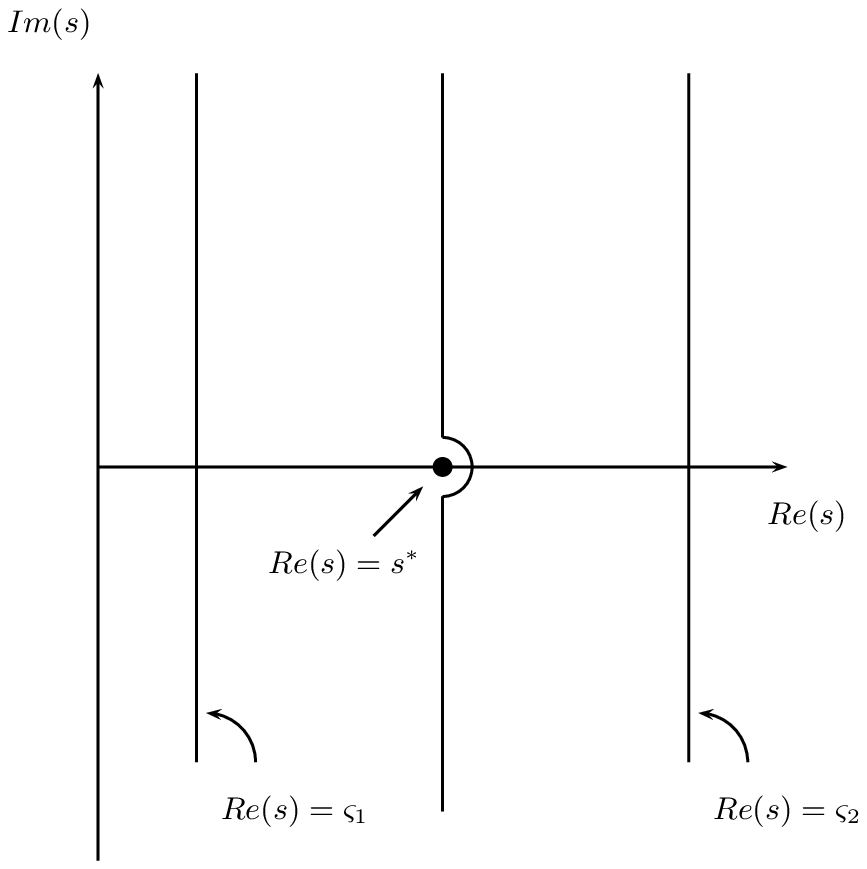}
\end{center} 
\caption{Integrating over a contour in the complex plane of $s$.}
\label{scomp}
\end{figure}

This point represents a precise value of $s$ for which, if we were
to perform the inverse Mellin transform over the vertical contour in
the complex plane of $s$ given by $Re(s)=s^*$, the perturbation
variables thus returned would generate finite scalars
$\delta\Psi_{0,4}$ on the past null cone of the origin, but
\emph{diverging} scalars on the Cauchy horizon. However, we maintain
this is not enough to conclude the Cauchy horizon is unstable, for
the following reasons:

\begin{enumerate} \item From our definition of $\bar{\sigma}$ \eqref{sigch},
for $\bs=2$, a real integer, we require $s=s^*\in\mathbb{R}$. Thus
there is only a single, isolated point in the $s$ complex plane at
which $s$ is such that $\bs=2$, and it lies on the real axis. When
performing the inverse Mellin transform, we must integrate over the
contour $Re(s)=s^*$ in the complex plane, where
$\varsigma_1<s^*<\varsigma_2$, as in Figure \ref{scomp}. Thus the
function $\gamma$ is recovered as \bes \gamma(y,r)=\f{1}{2\pi
i}\int_{s^*-i\infty}^{s^*+i\infty} r^s H(y;s) ds. \ees A well known
theorem in complex analysis (Cauchy's integral theorem), states that
we may continuously deform the contour of integration if the region
thus swept out does not contain any poles. From our solution for $H$
when $s=s^*$ (and thus $\bs=2$), we see that the integrand has no
poles due to the value of $y$. That the integrand has no poles due
to the value of $s$ is a technically very difficult question to
address fully, and is beyond the scope of this paper; however, some
analysis in this direction was carried out in Section 6 of
\cite{BandT2}, and there is evidence that no poles would be
encountered in the general solution for $H$.

Thus when performing the inverse Mellin transform we may avoid the
single, isolated point which makes the scalars diverge.

\item The diverging mode corresponds to a single isolated point, that is a set of zero
measure, in the $s$ plane. This is not generic in any sense; to
conclude an unstable Cauchy horizon we would be looking for an extended
region in the $s$ plane in which the modes diverge.

\item Note that $\bs=2$ lies between $\sigma=1$ and $\sigma=2$.
Thus the value $Re(s)=s^*$ means non-integer exponents in the
solution for $H$ near the past null cone; that is the solution is
non-analytic. From the point of view of critical collapse, we would
restrict our initial data to only consider analytic perturbations,
and thus would avoid the diverging mode altogether. However assuming analytic initial data is a very strong restriction that we do not feel is warranted in the present case.
\end{enumerate}

For these three reasons we conclude that the Cauchy horizon formed
in the collapse of self-similar timelike dust is stable at the level of the modes of the Mellin transform under even
parity perturbations with $l\geq 2$.

\subsection{SFSH}

On the second future similarity horizon, denoted $y=y_e$, we find
indicial exponents for the fourth order ODE in $H$ as
$\{0,1,2,\bar{\bar{\sigma}}\}$, where \be
\bar{\bs}=3-\left[\f{7-s-2\left(\f{y}{G}+\f{G}{y}\right)}{1+y
G''}\right]_{y=y_e}, \qquad \lim_{\mu\to 0}\bar{\bs}=-1.
\label{sigsfsh}\ee Again we find the scalars go like
$(y-y_e)^{\bar{\bs}-1}.$ We may write \bes \bar{\bs}-1=\alpha(\mu)
s+\beta(\mu). \ees Our initial data confined $Re(s)>0$, and it is
easily found that for $0<\mu<\mu_*$, both $\alpha(\mu)$ and
$\beta(\mu)$ are always negative. Thus the scalars
$\delta\Psi_{0,4}$ will diverge on the SFSH for \emph{all} values of
$Re(s)$ allowed by initial data (in contrast to the Cauchy horizon).

We conclude that the second future similarity horizon formed
in the collapse of self-similar timelike dust is unstable at the level of the modes of the Mellin transform under
even parity perturbations with $l\geq 2$.


\section{Conclusions}

We have examined the linear stability of the Cauchy horizon which
may form in the collapse of the self-similar
Lema\^{i}tre-Tolman-Bondi spacetime, due to even-parity
perturbations of the metric and matter tensors of multipole mode
$l\geq 2$. We have found that the Cauchy horizon is linearly stable at the level of the modes of the Mellin tranform of the perturbation variables. However, interestingly, the second future similarity horizon which follows
the Cauchy horizon is unstable.

A crucial question then is whether the same result applies to the full perturbation - that is, to the resummed Mellin modes. This is a highly nontrivial question. We note two possible approaches. One would be to try to determine the asymptotic behaviour of the solutions of the $s-$parametrised system of ODE's for large values of $|s|$, with a view to showing that the solutions fall-off at a rate that would guarantee existence of the contour integral giving the inverse Mellin transform. Another would be to employ the energy methods of \cite{brienscalar} to directly study the evolution of the perturbation without recourse to the Mellin transform. This approach is currently begin used by one of us (BCN) to study rigorously the even parity perturbations of self-similar Vaidya spacetime. In both cases, there are significant technical obstacles. For the asymptotic analysis, one would need global information about how the different independent solutions of the $s-$parametrized ODE's at different singular points are related to one another. For the energy methods, the transition from a scalar wave equation to a first order hyperbolic system gives rise to significant additional difficulties, principally in determining an appropriate energy functional. It is hoped that developing the appropriate `technology' for Vaidya spacetime (a simpler case) will yield results applicable to the present case.

However, we maintain that the results derived here are of physical relevance. We have found that a non-trivial necessary condition for linear stability is satisfied. Furthermore, and for example, an initially finite perturbation constructed from a finite number of Mellin modes will remain finite when it impinges on the Cauchy horizon.

Finally, we note that our results here mirror exactly those relating to the stability of the self-similar Vaidya spacetime previously studied by
the authors, namely that the naked singularity survives the
perturbation but only does so for a finite time. This adds further
weight to the observation of the authors in \cite{BandT2} that
perhaps a generic feature of naked singularities in self-similar
spacetimes is the linear stability of `fan'-type similarity horizons
(the Cauchy horizon) and instability of `splash'-type similarity
horizons (the SFSH), to use the terminology of Carr and Gundlach
\cite{CarrGund}.


\section*{Acknowledgments}

This research was funded by Enterprise Ireland grant SC/2001/199. TW would like to thank Dublin City University where the bulk of this work was carried out. Both authors would like to thank an anonymous referee for their helpful comments and queries.


\appendix

\section{Perturbation equations}

We give here the full set of perturbation equations for the gauge
invariant quantities defined in \S II.B. Note we only consider
multipole modes $l\geq 2$ and thus all equations are valid.

\begin{subequations} \label{perteqns}
\begin{eqnarray}
2v^C(k_{AB|C}-k_{CA|B}-k_{CB|A}+2\w{g}_{AB}k_{CD}^{\phantom{CD}|D})-2\w{g}_{AB}v^C
k_{D\phantom{D}|C}^{\phantom{D}D} +\w{g}_{AB}\!
\left(\f{l(l+1)}{r^2}+\!
\f{1}{2}(\w{G}_C^{\phantom{C}C}\!+\w{G}_a^{\phantom{a}a})\!+\!\w{\mathcal{R}}\right)\!
k_D^{\phantom{D}D}&&\nonumber\\
\!+2(v_Ak,_B+v_Bk,_A+k,_{A|B}) +\w{g}_{AB}(2 v^{C|D}+4v^C
v^D-\w{G}^{CD})k_{CD} -\w{g}_{AB}\left(2k,_C^{\phantom{,C}|C}+6v^C
k,_C-\f{(l-1)(l+2)}{r^2}k\right) &&\nonumber\\
 -\left(\f{l(l+1)}{r^2}+\w{G}_C^{\phantom{C}C}+\w{G}_a^{\phantom{a}a}+2\w{\mathcal{R}}\right)k_{AB}=-16 \pi T_{AB},& \label{tenspert}\\
-\left(k_{C\phantom{C}|D}^{\phantom{C}C\phantom{|D}|D}+\w{\mathcal{R}}
k_C^{\phantom{C}C}-\f{l(l+1)}{2r^2}k_C^{\phantom{C}C}\right)-(k,_C^{\phantom{C}|C}+2v^Ck,_C+\w{G}_a^{\phantom{a}a}k)\hspace{2cm}&&\nonumber\\
+\left(k_{CD}^{\phantom{CD}|C|D}+2v^Ck_{CD}^{\phantom{CD}|D}+2(v^{C|D}+v^C v^D)k_{CD}\right) =-16 \pi T^1, & \label{E10c}\\
k,_A-k_{AC}^{\phantom{AC}|C}+k_{C\phantom{C}|A}^{\phantom{C}C}-v_A k_C^{\phantom{C}C}=-16\pi T_A,&\\
k_A^{\phantom{A}A}=-16\pi T^2.& \label{tracek}
\end{eqnarray} \label{evenpar}
\end{subequations}

Here $\mathcal{R}$ is the Gaussian curvature of $\mc{M}^2$, the
manifold spanned by the time and radial coordinates, and thus equals
half the Ricci scalar of $\mc{M}^2$; also $\w{G}_{\mu\nu}$ is the
Einstein tensor of the background spacetime.

\section{Coefficients of the first order linear matrix equation}

We give here the coefficients of the first order linear system $Y'=M
Y$ of Section III:


\begin{align*} M_{11}=&\, G^4\left(l+l^2-2s^2 + 2sG' \right)  -
G^3y \left( l + l^2 + 2\left( s^2 + s-1 \right)  - 4\left(1 - l -
l^2 - \tfrac{s}{2} + s^2 \right) G'\right. \\ & \left. +
     \left( -1 + 8s \right) {G'}^2 \right)  + G^2y^2
   \left( l + l^2 + 2s + \left( -6 + 3l + 3l^2 + 4s + 2s^2 \right)G' \right. \\ & \left. +
     \left( -14 + 6l + 6l^2 + s - 2s^2 \right) {G'}^2
 + 2\left( -2 + 5s \right) {G'}^3 \right) +
  y^4G'\,\left( -2 + l + l^2 + \right. \\ & \left. \left( -4 + l + l^2 \right)G' + \left( -4 + l + l^2 \right){G'}^2 +
     \left( -5 + l + l^2 \right) {G'}^3 - {G'}^4 \right)\\ &  -
  Gy^3\left( -2 + l + l^2 + 2\left( -2 + l + l^2 + s \right) G' + 3\left( -3 + l + l^2 + s \right) {G'}^2 +
    \right. \\ & \left. \left( -14 + 4l + 4l^2 - s \right) {G'}^3 + 4\left( -1 + s \right){G'}^4
     \right) \\ & /G\left( 2G^2\left( 1 + s \right)  - 2G\left( 1 + s \right)yG' + y^2{G'}^2 \right)
  \left( G-yG'-y\right)\left(G-yG'+y  \right) \end{align*}\begin{align*}
  M_{12}=&\,G^4\left( l + l^2 - 2s^2 \right)  - 2G^3\left( -2 + 2l + 2l^2 + s - 2s^2 \right)yG' +
  y^4{G'}^2\left( -4 + l + l^2 \right. \\ & \left. + \left( -5 + l + l^2 \right){G'}^2 \right)   +
  Gy^3G'\left( -2\left( -2 + l + l^2 + s \right)  + \left( 14 - 4l - 4l^2 + s \right){G'}^2 \right) \\ & +
  G^2y^2\left( l + l^2 + 2s + \left( -14 + 6l + 6l^2 + s - 2s^2 \right){G'}^2
  \right)\\ & /G\left( 2G^2\left( 1 + s \right)  - 2G\left( 1 + s \right) yG' + y^2{G'}^2 \right)
   \left( G-yG'-y\right)\left(G-yG'+y\right)\left(yG'-G
   \right)\\
   M_{13}=&\,\left( G - yG' \right) \left( 2G^3sG' + \left( -2 + l + l^2 \right) y^3\left( 1 + {G'}^2 \right)  +
    G^2y\left( l + l^2 + 2\left( -1 + s + s^2 \right) \right. \right. \\ & \left. \left. - 5s{G'}^2 \right)  +
    Gy^2G'\left( -2\left( -2 + l + l^2 - s \right)  + 3s{G'}^2 \right)
    \right)\\ & /G\left( 2G^2\left( 1 + s \right)  - 2G\left( 1 + s \right) yG' + y^2{G'}^2 \right)
  \left( G-yG'-y\right)\left(G-yG'+y  \right) \\
  M_{14}=&\, -2{G'}^2\left( G^3\left( -1 + s \right)  - 2G^2\left( -2 + s \right)yG' + y^3G'\left( 2 + {G'}^2 \right)  +
    Gy^2\left( -3 - s + \right. \right. \\ & \left. \left. \left( -4 + s \right){G'}^2 \right)
    \right)\\ & /G\left( 2G^2\left( 1 + s \right)  - 2G\left( 1 + s \right)yG' + y^2{G'}^2 \right)
  \left( G-yG'-y\right)\left(G-yG'+y \right) \\ \bigskip \\
  M_{21}=&\, \left( G - yG' \right) \left( G - y\left( 1 + G' \right)  \right)
  \left( y^2\left( -2 + l + l^2 - 2G' \right) G'\left( 1 + G' \right)  +
    G^2\left( l + l^2 \right. \right. \\ & \left. \left. + 2s + 2sG' \right)  -
    Gy\left( -2 + l + l^2 + 2\left( -2 + l + l^2 + s \right)G' + \left( -3 + 2s \right){G'}^2 \right)
    \right)\\ & / G\left( G  - yG'+y \right)\left( 2G^2\left( 1 + s \right)  - 2G\left( 1 + s \right) yG' +
    y^2{G'}^2 \right) \\
  M_{22}=&\, -2G^4\left( l + l^2 - 2s^2 \right)  - y^4\left( 2\left( -4 + l + l^2 \right)  - G' \right) {G'}^3
   \left( 1 + G' \right)  + 2G^3y\left( l + l^2 + 2s \right. \\ & \left. + 2\left( -2 + 2l + 2l^2 + s - 2s^2 \right)G' -
     \left( 1 + s \right){G'}^2 \right)  + 2G^2y^2G'
   \left( 4 - 3l - 3l^2 - 4s \right. \\ & \left. + \left( 13 - 6l - 6l^2 - 2s + 2s^2 \right)G' +
     2\left( 1 + s \right){G'}^2 \right) \\ & + Gy^3{G'}^2
   \left( 6l + 6l^2 + 4\left( -4 + s \right)  + \left( -26 + 8l + 8l^2 \right)G' -
     \left( 3 + 2s \right) {G'}^2 \right) \\ & /2G\left( 2G^2\left( 1 + s \right)  - 2G\left( 1 + s \right)yG' + y^2{G'}^2 \right)
  \left( G-yG'\right)\left(G-yG'+y  \right) \\  M_{23}=&\, {\left( G - y\,G' \right) }^2\,\left( -2\,G^2\,s\,\left( 1 + s - G' \right)  -
    \left( -2 + l + l^2 \right) \,y^2\,\left( 1 + G' \right) \right. \\ & \left. + G\,y\,\left( -2 + l + l^2 - 2\,s\,G' - 3\,s\,{G'}^2 \right)
    \right) \\ & /G\,\left( G  - y\,G'+y \right) \,\left( 2\,G^2\,\left( 1 + s \right)  - 2\,G\,\left( 1 + s \right)
     \,y\,G' + y^2\,{G'}^2 \right) \\
   M_{24}=&\, -2\,{G'}^2\,\left( -G + y\,G' \right) \,\left( -\left( G^2\,\left( -1 + s \right)  \right)  + y^2\,G'\,\left( 2 + G' \right) \right. \\ & \left.  +
    G\,y\,\left( -3 - s + \left( -3 + s \right) \,G' \right)
    \right)\\ & / G\,\left( G  - y\,G'+ y \right) \,\left( 2\,G^2\,\left( 1 + s \right)  - 2\,G\,\left( 1 + s \right) \,y\,G' + y^2\,{G'}^2
    \right)\\ M_{31}=&\,\left( -\left( y^2\,\left( -2 + l + l^2 - 2\,G' \right) \,G'\,\left( 1 + G' \right)  \right)  -
  G^2\,\left( l + l^2 + 2\,s + 2\,s\,G' \right)\right. \\ & \left. + G\,y\,
   \left( -2 + l + l^2 + 2\,\left( -2 + l + l^2 + s \right) \,G' + \left( -3 + 2\,s \right) \,{G'}^2
   \right) \right) \\ & / G\,\left( 2\,G^2\,\left( 1 + s \right)  - 2\,G\,\left( 1 + s \right) \,y\,G' + y^2\,{G'}^2
   \right)\\
  M_{32}=&\,\left(G\,\left( l + l^2 + 2\,s \right)  - \left(l^2+l-4 \right)
  \,y\,G'\right)/G\left( 2\,G^2\,\left( 1 + s \right)  - 2\,G\,\left( 1 + s \right) \,y\,G' + y^2\,{G'}^2
  \right)\\
  M_{33}=&\, \left(-2\,G^2\,s\,G' + \left( -2 + l + l^2 \right) \,y^2\,G' - G\,y\,\left( -2 + l + l^2 - 3\,s\,{G'}^2
  \right)\right) \\ & / G\,\left( 2\,G^2\,\left( 1 + s \right)  - 2\,G\,\left( 1 + s \right) \,y\,G' + y^2\,{G'}^2
  \right)\\
  M_{34}=&\,\left( 2\,{G'}^2\,\left( -\left( G\,\left( 3 + s \right)  \right)  + 2\,y\,G'
  \right)\right) /G\,\left( 2\,G^2\,\left( 1 + s \right)  - 2\,G\,\left( 1 + s \right) \,y\,G' + y^2\,{G'}^2
  \right)\\ \bigskip \\
  M_{41}=&\, -1/2 \qquad M_{42}=0 \qquad M_{43}=0 \qquad
  M_{44}=0.
\end{align*}

\bibliographystyle{plain}
\bibliography{LTB}

\begin{thebibliography}{10}

\bibitem{CarrGund}
B.J. Carr and C.~Gundlach.
\newblock Spacetime structure of self-similar spherically symmetric perfect
  fluid solutions.
\newblock {\em Physical Review D}, (67):024035, 2003.

\bibitem{chandhartle}
S.~Chandrasekher and J.~Hartle.
\newblock On crossing the {C}auchy horizon of a {R}eissner-{N}ordstr\o m black
  hole.
\newblock {\em Proceedings of the Royal Society of London}, A(384):301, 1982.

\bibitem{christi}
D.~Christodoulou.
\newblock Examples of naked singularity formation in the gravitational collapse
  of a scalar field.
\newblock {\em Annals of Mathematics}, (140):607--653, 1994.

\bibitem{dafermos1}
M.~Dafermos.
\newblock Stability and instability of of the {C}auchy horizon for the
  spherically symmetric {E}instein-{M}axwell-scalar field equations.
\newblock {\em Annals of Mathematics}, (158):875--928, 2003.

\bibitem{dafermos2}
M.~Dafermos.
\newblock The interior of charged black holes and the problem of uniqueness in
  general relativity.
\newblock {\em Comm. Pure Appl. Math.}, (58):445--504, 2005.

\bibitem{Frolov2}
A.~Frolov.
\newblock Perturbations and critical behaviour in the self-similar
  gravitational collapse of a massless scalar field.
\newblock {\em Physical Review D}, (56):6433--6438, 1997.

\bibitem{Frolov1}
A.~Frolov.
\newblock Critical collapse beyond spherical symmetry: general perturbations of
  the {R}oberts solution.
\newblock {\em Physical Review D}, (59):104011, 1999.

\bibitem{GS1}
U.H. Gerlach and U.K. Sengupta.
\newblock Gauge-invariant perturbations on most general spherically symmetric
  spacetimes.
\newblock {\em Physical Review D}, 19(8):2268--2272, April 1979.

\bibitem{GS2}
U.H. Gerlach and U.K. Sengupta.
\newblock Gauge-invariant coupled gravitational, acoustical, and
  electromagnetic modes on most general spherical spacetimes.
\newblock {\em Physical Review D}, 22(6):1300--1312, Sept. 1980.

\bibitem{jg}
J.B. Griffiths.
\newblock The stability of {K}illing-{C}auchy horizons in colliding plane wave
  space-times.
\newblock {\em General Relativity and Gravitation}, (37):1119--1128, 2005.

\bibitem{gundfluid}
C.~Gundlach.
\newblock Critical gravitational collapse of a perfect fluid: nonspherical
  perturbations.
\newblock {\em Physical Review D}, (65):084021, 2002.

\bibitem{gundmartin}
C.~Gundlach and J.M. Mart\'{\i}n-Garc\'{\i}a.
\newblock All nonspherical perturbations of the {C}hoptuik spacetime decay.
\newblock {\em Physical Review D}, (59):064031, 1999.

\bibitem{livrev}
C.~Gundlach and J.M. Mart\'in-Garc\'ia.
\newblock Critical phenomena in gravitational collapse.
\newblock {\em Living Reviews in Relativity}, 2(4), 1999.
\newblock URL (cited on 18/03/2009) www.livingreviews.org/lrr-1999-4.

\bibitem{Gund}
C.~Gundlach and J.M. Mart\'{\i}n-Garc\'{\i}a.
\newblock Gauge-invariant and coordinate-independent perturbations of stellar
  collapse {I}: the interior.
\newblock {\em Physical Review D}, 61:084024, 2000.

\bibitem{haradaodddust}
T.~Harada, H.~Iguchi, and K.~Nakao.
\newblock Gravitational radiation from a naked singularity {I}: odd-parity
  perturbation.
\newblock {\em Progress of Theoretical Physics}, (101):1235--1252, 1999.

\bibitem{haradaevendust}
T.~Harada, H.~Iguchi, and K.~Nakao.
\newblock Gravitational radiation from a naked singularity {II}: even-parity
  perturbation.
\newblock {\em Progress of Theoretical Physics}, (103):53--72, 2000.

\bibitem{harwave}
T.~Harada, H.~Iguchi, and M.~Shibata.
\newblock Computing gravitational waves from slightly nonspherical collapse to
  a black hole: odd-parity perturbation.
\newblock {\em Physical Review D}, (68):024002, 2003.

\bibitem{joshi}
P.S. Joshi.
\newblock {\em Global aspects in gravitation and cosmology}.
\newblock Clarendon Press, Oxford, 1993.

\bibitem{LittlefieldDesai}
D.L. Littlefield and P.V. Desai.
\newblock Frobenius analysis of higher order equations: incipient buoyant
  thermal convection.
\newblock {\em SIAM Journal on Applied Mathematics}, 50(6):1752--1763, Dec.
  1990.

\bibitem{Rezzolla}
A.~Nagar and L.~Rezzolla.
\newblock Gauge-invariant non-spherical metric perturbations of {S}chwarzschild
  spacetime.
\newblock {\em Classical and Quantum Gravity}, (22):R167, 2005.

\bibitem{brienscs}
B.C. Nolan.
\newblock Dynamical extensions for shell-crossing singularities.
\newblock {\em Classical and Quantum Gravity}, (20):575--586, 2003.

\bibitem{brienscalar}
B.C. Nolan.
\newblock Bounds for scalar waves on self-similar naked singularity
  backgrounds.
\newblock {\em Classical and Quantum Gravity}, (23):4523--38, 2006.

\bibitem{brienphillipe}
B.C. Nolan and F.~Mena.
\newblock Geometry and topology of singularities in spherical dust collapse.
\newblock {\em Classical and Quantum Gravity}, (19):2587--2605, 2002.

\bibitem{BandT1}
B.C. Nolan and T.J. Waters.
\newblock Cauchy horizon stability in self-similar collapse: scalar radiation.
\newblock {\em Physical Review D}, (66):104012, 2002.

\bibitem{BandT2}
B.C. Nolan and T.J. Waters.
\newblock Even perturbations of self-similar {V}aidya spacetime.
\newblock {\em Physical Review D}, (71):104030, 2005.

\bibitem{oripiran}
A.~Ori and T.~Piran.
\newblock Naked singularities and other features in self-similar
  general-relativistic collapse.
\newblock {\em Physical Review D}, 42(4):1068--1090, 1990.

\bibitem{poisson}
E.~Poisson and W.~Israel.
\newblock Internal structure of black holes.
\newblock {\em Physical Review D}, (41):1796--1809, 1990.

\bibitem{Sarbach}
O.~Sarbach and M.~Tiglio.
\newblock Gauge invariant perturbations of {S}chwarzschild black holes in
  horizon-penetrating coordinates.
\newblock {\em Physical Review D}, (64):084016, 2001.

\bibitem{Wald}
R.M. Wald.
\newblock {\em General Relativity}.
\newblock University of Chicago press, Chicago, 1984.

\bibitem{Wang}
Z.X. Wang and D.R. Guo.
\newblock {\em Special functions}.
\newblock World Scientific, Singapore, 1989.

\end{thebibliography}

\end{document}